\documentclass[a4paper]{jpconf}

\usepackage{graphicx,color}

\usepackage{amsmath,amsfonts,amssymb,amsthm}

\newtheorem{teo}{Theorem}

\renewcommand{\theenumi}{\roman{enumi}}

\begin{document}

\title[]{Extensions of Natural Hamiltonians}

\author{G Rastelli}
\address{Independent researcher, cna Ortolano 7, Ronsecco, Italia} 
\ead{giorast.giorast@alice.it}

\begin{abstract} Given an n-dimensional natural Hamiltonian L on a Riemannian or pseudo-Riemannian manifold, we call "extension" of L the n+1 dimensional Hamiltonian $H=\frac 12 p_u^2+\alpha(u)L+\beta(u)$ with new canonically conjugated coordinates $(u,p_u)$. For suitable L, the functions $\alpha$ and $\beta$ can be chosen depending on any natural number m such that H admits an extra polynomial first integral in the momenta of degree m, explicitly  determined in the form of the m-th power of a differential operator applied to a certain function of coordinates and momenta. In particular, if L is maximally superintegrable (MS) then H is MS also. Therefore, the extension procedure allows the creation of new superintegrable systems from old ones. For m=2,  the extra first integral generated by the extension procedure determines a second-order symmetry operator of a  Laplace-Beltrami quantization of H, modified by taking in account the curvature of the configuration manifold. The extension procedure can be applied to several Hamiltonian systems, including the three-body Calogero and Wolfes systems (without harmonic term),  the Tremblay-Turbiner-Winternitz system and n-dimensional anisotropic harmonic oscillators. We propose here a short review of the known results of the theory and some previews of new ones. 
\end{abstract}

\section{Introduction} 

Natural Hamiltonian systems are the mathematical models of those physical systems for which the energy is constant, for example harmonic oscillators or the Kepler system. Often, as in the previous two examples, more quantities are constants of the motion (or {\it first integrals}):  angular momentum, Laplace-Runge-Lentz vector, etc. Usually, these constants are expressed by quadratic polynomials in the momenta or, for quantum systems, by second-order differential operators. Hamiltonian systems with constants of the motion of degree higher than two are less common, nevertheless, some of them are of great  interest, as for instance the three-body Jacobi-Calogero  and  Wolfes  systems. These systems represent the dynamics of three point-masses on a line under forces determined by the potential functions 
\begin{eqnarray*}
V_{JC}=k\left((x-y)^{-2}+(x-z)^{-2}+(y-z)^{-2}\right),\\ V_W=k\left((x+y-2z)^{-2}+(x+z-2y)^{-2}+(y+z-2x)^{-2}\right),
\end{eqnarray*}
respectively (we do not consider here the harmonic oscillator terms) and they have essentially the same dynamics \cite{CDR1}. Both the resulting natural Hamiltonians in $\mathbb E^3$ admit one linear and one  quadratic in the momenta constants of the motion, making the systems Liouville-integrable and solvable by separation of variables (see \cite{CDR1} and references therein). Other two independent  constants of the motion do exist, one quadratic, due to the multiseparability of the Hamiltonian, and one cubic. The systems are then maximally superintegrable (MS), having a number of functionally independent constants of the motion equal to twice the degrees of freedom, minus one (for quantum systems, the same number of algebraically independent symmetry operators). MS systems are of the greatest importance in mathematical physics, harmonic oscillators and Kepler are MS and this  makes them to satisfy  Bertrand's theorem. Indeed, maximal superintegrability manifests itself, for classical systems, in the fact that all finite orbits of MS systems are closed while, for quantum systems, in the fact that  the energy levels are totally degenerate \cite{JH}.
In recent years, several techniques made possible the construction of classical and quantum Hamiltonian systems, MS and not, with first integrals of arbitrarily high degree \cite{CDR1,TTW,KMK,POL,MPY} whose study, still in development, produced remarkable results in special functions,  quantum algebras, canonical quantization theories \cite{KMK,KMK2,Ri,Ra2,Lev}.  In this note it is shortly introduced the work on the topic done by Claudia Chanu, Luca Degiovanni and the author (in short CDR) in several joint articles. In few words, the extension procedure  (Theorem \ref{Teo0}) adds one degree of freedom to some suitable Hamiltonian $L$  in such a way an extra non trivial first integral, polynomial of degree $m\in \mathbb N$, of the new Hamiltonian do exist.  

\section*{Main results}
The following Theorem, stated in \cite{sigma11}, defines and characterizes what we intend for "extensions" in the particular case of natural Hamiltonians on cotangent bundles of Riemannian manifolds; for a more general definition, see \cite{CDRG}.
Given an $n$-dimensional natural Hamiltonian $L$ on the cotangent bundle of a (pseudo)-Riemannian manifold $Q$, let be
\begin{equation}\label{HamExt}
H=\frac{1}{2} p_u^2+\alpha(u)L + \beta(u)
\end{equation}
and
\begin{equation}\label{U}
U=p_u+ \gamma(u) X_L,
\end{equation}
where $X_L$ is the Hamiltonian vector field of $L$, then

\teo \label{Teo0}
Let  $Q$ be a $n$-dimensional (pseudo-)Riemannian manifold with metric tensor $\mathbf g$. 
The natural Hamiltonian $L=\frac{1}{2} g^{ij}p_ip_j+V(q^i)$ on $M=T^*Q$ with canonical coordinates $(p_i,q^i)$ admits an extension $H$ in the form (\ref{HamExt}) 
with a first integral $F=U^m(G)$ with $U$ given by (\ref{U}) and $G(q^i)$,  if and only if the following conditions hold:
\begin{enumerate}

\item
the functions $G$ and $V$ satisfy
\begin{equation}\label{HessTeo} 
 \mathbf{H}(G)\,+\,mc\, \mathbf{g}G=\mathbf 0, \quad m\in \mathbb N-\{0\}, c\in \mathbb R,
\end{equation}
\begin{equation}
\label{VTeo} \nabla V \cdot \nabla G-2m(cV+L_0)G=0,\quad L_0\in \mathbb R,
\end{equation}
where $\mathbf H(G)_{ij}=\nabla_i\nabla_jG$ is the Hessian tensor of $G$.
\item for $c=0$ the extended Hamiltonian $H$ and the first integral $U^mG$ are
\begin{equation}\label{Ext0}
H=\frac{1}{2}p_u^2+mA(L+V_0)+mL_0A^2(u+u_0)^2, \quad U^mG=\left(p_u-A(u+u_0)X_L\right)^mG, 
\end{equation}
for $c\neq 0$ the extended Hamiltonian $H$ and the first integral $U^mG$ are
\begin{equation}\label{Extk}
H=\frac{1}{2}p_u^2+\frac{m(cL+L_0)}{S^2_\kappa(cu+u_0)}+W_0,\quad U^mG=
\left(p_u+\dfrac {C_\kappa (cu+u_0)}{S_\kappa (cu+u_0)}X_L\right)^mG, 
\end{equation}
with $\kappa$, $u_0,V_0, W_0, A\in \mathbb R$, $A\neq0$ and
\begin{equation*}
S_\kappa(x)=\left\{\begin{array}{ll}
\frac{\sin\sqrt{\kappa}x}{\sqrt{\kappa}} & \kappa>0, \\
x & \kappa=0, \\
\frac{\sinh\sqrt{|\kappa|}x}{\sqrt{|\kappa|}} & \kappa<0.
\end{array}\right. 
\qquad C_\kappa(x)=\left\{\begin{array}{ll}
{\cos\sqrt{\kappa}x} & \kappa>0, \\
1 & \kappa=0, \\
{\cosh\sqrt{|\kappa|}x} & \kappa<0.
\end{array}\right.
\end{equation*}

\end{enumerate}

\rm 

 In \cite{sigma11} it is proved that $U^m(G)$ is functionally independent from $H$, $L$ and any  other first  integral of $L$ in $T^*Q$.  The integrability conditions of  (\ref{HessTeo}) are discussed in \cite{sigma11} and it is found that their complete integrability requires $mc=\kappa$, where $\kappa$ is the constant curvature of $Q$. Then, the function $G(q^i)$ can depend linearly on up to $n+1$ parameters and the maximal number of parameters is attained on constant curvature manifolds only. However, non complete solutions can be found in non-constant curvature manifolds  (CDR to appear). From equation (\ref{VTeo}), the expressions of the admissible potentials $V$ can  be computed. Several examples are given in \cite{sigma11,CDRG,SE}. 

The particular form of $U$ makes possible to explicit any $U^m(G)$ by expanding the $m$-th power of a binomial, obtaining \cite{CDRG} 
\begin{equation}\label{EE}
U^mG=P_mG+D_mX_LG,
\end{equation}
with
$$
P_m=\sum_{k=0}^{[m/2]}{\left( \begin{matrix} m \cr 2k \end{matrix} \right) \gamma^{2k}p_u^{m-2k}\left(-2m(cL+L_0)\right)^k},
$$
$$
D_m=\sum_{k=0}^{[(m-1)/2]}{\left( \begin{matrix} m \cr 2k+1 \end{matrix} \right) \gamma^{2k+1}p_u^{m-2k-1}\left(-2m(cL+L_0)\right)^k}, \quad m>1,
$$
where $[\cdot]$ denotes the integer part and $D_1=\gamma$. We remark that first integrals of high degree obtained in other ways  than by the extension procedure \cite{KMK,MPY} can be explicitly expressed only thanks to the fact that the dynamical equations  are in these cases separated in some coordinate system.

As a first example of the extension procedure we consider the one-dimensional Hamiltonian \cite{sigma11}
\begin{equation*}
L=\frac 12 p_v^2+V(v).
\end{equation*}
The geodesic term of the extended Hamiltonian $H$ is
\begin{equation*}
\frac{1}{2}\left(p_u^2+\frac{mc}{S_\kappa^2(cu+u_0)}p_v^2\right),
\end{equation*}
where $\kappa$ is here the constant curvature of the extended configuration manifold.
The solutions of equations (\ref{HessTeo}) and (\ref{VTeo}) are
\begin{eqnarray*}
G(v) = b_1 S_{mc}(v+v_0), \quad
V(v) =b_2\frac{1}{C_{mc}^2(v+v_0)},
\end{eqnarray*}
where $b_i\in \mathbb R$.
When $c=1$ and $\kappa=0,1,-1$, the configuration manifold of $H$ is   the Euclidean plane, the Sphere $\mathbb S_2$ and the pseudosphere $\mathbb H_2$ respectively, while  for $c=-1$ and $\kappa=0,1,-1$, the Minkowski plane, the deSitter and anti-deSitter  manifolds,  respectively. After a rescaling of the coordinate $v$, the parameter $m$ in $H$ passes into $L$ and
\begin{equation*}
V(v) =b_2\frac{1}{C_{mc}^2(mv+v_0)}.
\end{equation*}
This makes evident that the extension procedure introduces some discrete symmetry into $H$, in this case a dihedral symmetry of order $2m$, somehow connected with the extra first integral $U^m(G)$. In the Euclidean plane (i.e. $\kappa=0$) with $c=m^{-1}$ and $m=3$, $V$ is associated with the Jacobi-Calogero potential or, equivalently, with the Wolfes potential \cite{CDR1,POL}. Indeed, in cylindrical coordinates of $\mathbb E^3$, $(u,v,w)$, with axis $w$ parallel to $(1,1,1)$ w.r. to Cartesian coordinates $(x,y,z)$, we have
\begin{equation*}
V_{JC}=\frac k{u^2\sin ^2(3v)}, \quad V_{W}=\frac k{u^2\cos ^2(3v)}.
\end{equation*}

The procedure of extension provides two new functionally independent first integrals to the extended Hamiltonian $H$: $H$ itself and $U^mG$. This fact is particularly relevant when the Hamiltonian $L$ is MS. In this case, $H$ is MS too, admitting $2n+1=2(n+1)-1$ functionally independent first integrals. In \cite{SE} this property of extended Hamiltonians is studied in several cases. For example, let us consider 

\begin{equation*}
L=\frac 12 p_1^2+\frac \zeta{S_\chi^2(x_1)}\left(\frac12 p_2^2+F(x_2)\right), \quad \chi, \zeta \in \mathbb R,
\end{equation*}
that is a  particular case of the generalized Tremblay-Turbiner-Winternitz system (TTW) \cite{TTW,KMK,MPY} for $\zeta=1$ and
\begin{equation*}\label{TTW}
F(x_2)=\frac{\alpha_1}{\cos^2\lambda x_2}+\frac{\alpha_2}{\sin^2\lambda x_2},
\end{equation*}

defined on constant-curvature manifolds of curvature $\chi$. This system is MS for any rational parameter $\lambda$ and admits polynomial first integrals of degree related to $\lambda$ \cite{TTW,KMK,MPY}.  In \cite{SE} it is shown that  $L$ always admits extensions of the form
\begin{equation*}
H=\frac 12p_u^2+\frac \chi{S_\kappa^2(\frac \chi m u)}L, 
\end{equation*}
with $G=a_0C_\chi(x_1)+\left(  a_1S_\zeta(x_2)+a_2C_\zeta(x_2)\right)S_\chi(x_1)$, creating in this way  new MS systems. Similarly, harmonic oscillators in $\mathbb E^n$, isotropic or not, can be extended into harmonic oscillators in $\mathbb E^{n+1}$ \cite{SE}. The assumption $G(q^i)$ can be generalized to $G(q^i,p_i)$. This leads to important generalizations, introduced in \cite{CDRG}, that will be developed in a  paper in preparation (CDR). The extension procedure can be in this case applied with $m$ substituted by any positive rational $\mu$ after a suitable definition of $U^\mu G$, so that the generalized TTW system of above, with $\alpha_1 \alpha_2=0$, can be written as an extension for any rational $\lambda$.

To classical extended Hamiltonians and their first integrals can be associated quantum Hamiltonians and symmetry operators by some procedure of quantization, usually in form of Laplace-Beltrami operators. When the curvature of $Q$ is not constant, the quantization requires additional terms (quantum corrections) in order to keep integrability or superintegrability . The quantum correction is then determined by the scalar curvature and by the Weyl tensor \cite{KMK2,Ri}. At least in the case $m=2$, the simultaneous quantization of $H$ and $U^2G$ is possible, allowing the preservation of maximal superintegrability of MS classical extended systems to the quantum limit. This will be shown in a paper in preparation (CDR).

\end{document}